\documentclass[aps,twocolumn,amsmath,amssymb,floatfix]{revtex4-1}
\usepackage{graphicx}
\usepackage{dcolumn}
\usepackage{bm}
\usepackage{amsfonts}
\usepackage{graphicx}
\usepackage{bm}
\usepackage{amsmath}
\usepackage{amssymb}

\begin{document}
\title{Polarizability and Screening in Chiral Multilayer Graphene}
\author{Hongki Min$^{1}$}
\author{E. H. Hwang$^{2}$}
\author{S. Das Sarma$^{2}$}
\affiliation{
$^{1}$Department of Physics and Astronomy, Seoul National University, Seoul 151-747, Korea\\
$^{2}$Condensed Matter Theory Center, Department of Physics, University of Maryland, College Park, Maryland 20742, USA  
}
\date{\today}

\begin{abstract}
We calculate the static polarizability of multilayer graphene 
and study the effect of stacking arrangement, carrier density, and on-site energy difference on graphene screening properties.
At low densities, the energy spectrum of multilayer graphene is described by a set of chiral two-dimensional electron systems, and the associated chiral nature determines the screening properties of multilayer graphene, showing very different behavior depending on whether the chirality index is even or odd.
As density increases, the energy spectrum follows that of the monolayer graphene and thus the polarizability approaches that of monolayer graphene. 
The qualitative dependence of graphene polarizability on chirality and
layering indicates the possibility of distinct graphene quantum phases as a function of the chirality index.

\end{abstract}

\maketitle

Since the discovery of graphene in 2004, the experimental fabrication of gated graphene has
been a subject of great interest\cite{review}. Recently,
multilayer graphene has attracted considerable attention\cite{yacoby2011}
for its fundamental properties and chiral electronic structure, and
for future applications due to the presence of an additional layer degree of
freedom. 
Interestingly, the electronic structure of multilayer graphene is sensitive to its stacking sequences and it could support various chiral two-dimensional electron systems (C2DESs) with different chirality indices, depending on the stacking sequences\cite{spectrum,min2008}.
Thus multilayer graphene is not a simple extension of monolayer or bilayer graphene and could open the possibility of engineering electronic properties by tuning the stacking arrangement.   

The optical properties of multilayer graphene have been studied
experimentally\cite{optical_experiment} and
theoretically\cite{optical_theory}, showing that characteristic peak
positions in the optical conductivity depend on stacking
sequences. The transport properties of multilayer graphene depend
sensitively on the number of layers and stacking sequences in the system\cite{transport_theory}. 
Quantum Hall effects in trilayer graphene
show different quantization rules compared with monolayer and bilayer
graphene\cite{qhe}. Electron-electron interactions also play an
important role in multilayer graphene, especially in periodic ABC
stacking compared with other stacking sequences due to the appearance
of relatively flat bands near the Fermi energy, indicating the
possibility of interaction induced ordered states\cite{fan2010} which
could be very different from those studied in monolayer and bilayer
graphene.  

As the number of graphene layers increases, the effects of screening
become more and more important due to the enhancement of the
electronic density of states (DOS), which determines the fundamental
properties such as transport, optical properties, phonon
dispersion, and the Ruderman-Kittel-Kasuya-Yosida (RKKY) interaction. 
The energy band structure of
multilayer graphene is very sensitive to its stacking sequence,
leading to its screening properties depending strongly on the stacking
arrangements. There has been extensive theoretical activity on Coulomb
screening in monolayer\cite{screening_monolayer} and bilayer
graphene\cite{screening_bilayer}, 
but to our knowledge a systematic study of screening in multilayer graphene is lacking.

In this paper,
we calculate the static polarizability of multilayer
graphene and systematically study the effect of stacking arrangement,
carrier density, and on-site energy difference on its screening
properties. We first study the screening properties of a C2DES with an arbitrary chirality
$J$, demonstrating that the polarizability and the associated screening
behavior strongly depend on $J$ due to the enhanced (suppressed)
backscattering for even (odd) $J$. Next we calculate the screening
properties of multilayer graphene, and then compare and analyze the
results using those of C2DES. 
At low densities, the energy spectrum of multilayer
graphene is described by a set of C2DESs\cite{min2008}, and screening
is determined by its chiral nature. As density increases, the energy
spectrum eventually follows that of monolayer graphene and thus the
polarizability approaches that of monolayer graphene in the high
density limit. We also show that the on-site energy difference between
the sublattices or layers, which could induce a band gap, can enhance
or suppress the backscattering depending on the chiral nature of
multilayer graphene. 

The static polarizability is defined by
\begin{equation}
\Pi(\bm{q})=-g\sum_{\lambda,\lambda'}\int {d^2 k \over (2\pi)^2} {f_{\lambda,\bm{k}}-f_{\lambda',\bm{k}'} \over \varepsilon_{\lambda,\bm{k}}-\varepsilon_{\lambda',\bm{k}'}} F_{\lambda,\lambda'}(\bm{k},\bm{k}'),
\label{eq:bare_polarization}
\end{equation}
where $g=g_{\rm s} g_{\rm v}$ is the total degeneracy factor 
($g_{\rm s}=g_{\rm v}=2$ are spin and valley degeneracy factors,
respectively), $\varepsilon_{\lambda,\bm{k}}$ and $f_{\lambda,\bm{k}}$
are the eigenenergy and Fermi function for the band index $\lambda$
and wave vector $\bm{k}$, respectively,
$F_{\lambda,\lambda'}(\bm{k},\bm{k}')$ is the square of the
wave-function overlap between $\left|\lambda,\bm{k}\right>$ and
$\left|\lambda',\bm{k}'\right>$ states, and $\bm{k}'=\bm{k+q}$.

To understand the screening properties of multilayer graphene,
we first consider a two-band pseudospin Hamiltonian which describes two-dimensional (2D)
chiral quasiparticles\cite{min2008}. The pseudospin Hamiltonian with
the chirality index $J$ is of the form 
\begin{equation}
H_J(\bm{k})=t_{\perp}\left(
\begin{array}{cc}
0 & \left({\hbar v_0 k_{-}\over t_{\perp}}\right)^J \\
\left({\hbar v_0 k_{+}\over t_{\perp}}\right)^J & 0 \\
\end{array}
\right),
\end{equation}
where $k_{\pm}=k_x\pm i k_y$, $v_0$ is the effective in-plane Fermi velocity, and $t_{\perp}$ is the nearest-neighbor interlayer hopping.
The Hamiltonian has an energy spectrum given by
$\varepsilon_{s,\bm{k}}=s t_{\perp} \left(\hbar v_0 |\bm{k}| \over
t_{\perp}\right)^J$ and the corresponding eigenfunctions are
$\left|s,\bm{k}\right>={1 \over \sqrt{2}}\left(s,e^{i
  J\phi_{\bm{k}}}\right)$, where $\phi_{\bm{k}}=\tan^{-1}(k_y/k_x)$
and $s=\pm 1$ for positive (negative) energy states,
respectively. Note that for C2DESs, $F_{s,s'}(\bm{k},\bm{k}')={1\over
  2}\left[1+s s'\cos J(\phi_{\bm{k}}-\phi_{\bm{k}'})\right]$, thus the
backscattering with $\phi_{\bm{k}}-\phi_{\bm{k}'}=\pm \pi$ between the
same bands $s=s'$ is enhanced (suppressed) for even (odd) $J$ due to
the chiral nature of the electronic structure. This has important
consequences as the scattering properties of odd and even $J$ turn out to be
qualitatively different. 

To investigate the polarizability we rewrite Eq.~(\ref{eq:bare_polarization}) as $\Pi(q)=\Pi_{\rm intra}(q) + \Pi_{\rm inter} (q)$, where 
the intraband polarizability $\Pi_{\rm intra}$ is defined by the intraband transition terms $\lambda=\lambda'$ in Eq.~(\ref{eq:bare_polarization}), and the interband polarizability $\Pi_{\rm inter}$ with $\lambda \neq \lambda'$ in Eq.~(\ref{eq:bare_polarization}) which is induced by the virtual interband transitions.

For the undoped intrinsic case (with the carrier density $n$ being zero), we find $\Pi_{\rm intra}(q)=0$ and 
$\Pi_{\rm inter}(q) = N(q) I(J)$, where $N(q)={g q^{2-J} \over 2\pi J t_{\perp}(\hbar v_0/t_{\perp})^J}$ and $I(J)$ is a constant depending only on the chirality index $J$, given by
\begin{equation}
I(J) = J\int_0^{\infty}xdx\int_0^{2\pi} \frac{d\phi}{2\pi} \frac{1-\cos (J\theta)}{x^J+(x')^{J}},
\end{equation}
where $x'=\sqrt{1+2x\cos\phi+x^2}$, and $\cos\theta = (x+\cos \phi)/x'$.
For example, we have $I(1)=\pi/8$ and $I(2) = \log 4$. 
Note that the screened Coulomb potential is given by $V_{\rm sc}(q) = 2 \pi e^2/\left[\kappa \left(q+q_{\rm s}(q)\right)\right]$, where $\kappa$ is the background dielectric constant and $q_{\rm s}(q) = (2\pi e^2/\kappa) \Pi(q)$ is the screening wave vector.
For the intrinsic case, 
$q_{\rm s}(q) \propto q^{2-J}$, 
and thus in the long wavelength limit, the screened Coulomb potential
behaves as $V_{\rm sc}(q) \propto q^{J-2}$ and for $J\ge 3$ it
vanishes as $q \rightarrow 0$, 
whereas $V_{\rm sc}(q) \propto 1/q$ as
$q \rightarrow \infty$ for all $J$. Thus, the long wavelength Coulomb
interaction is completely screened for $J\ge 3$.  

\begin{figure}
\includegraphics[width=1.0\linewidth,height=4.5in]{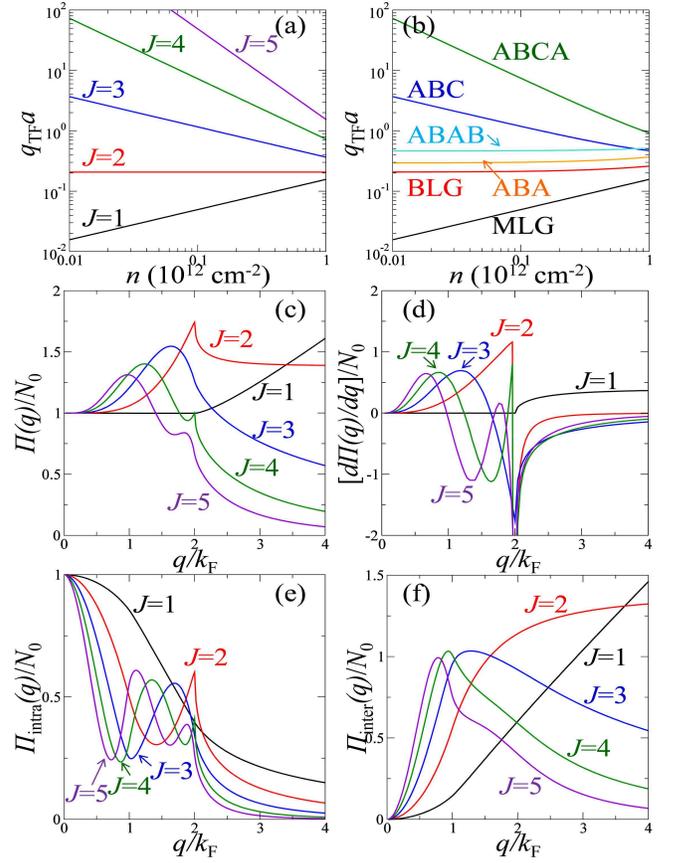}
\caption{
(a) The Thomas-Fermi wave vector, $q_{\rm TF} \propto N_0 =\Pi(q=0)$, as a function of density $n$ with $\kappa=2.5$, where $a=0.246$ nm is the lattice constant of graphene. (b) Same as (a) for multilayer graphene with several stacking sequences, where MLG and BLG represent monolayer and bilayer graphene, respectively. 
(c) and (d) show the total static polarizability $\Pi(q)$ and
its first derivative $d\Pi(q)/dq$, respectively, as a function of
wave vector for C2DES with chiralities $J=1,2,3,4,5$. 
Note that the first derivative
  at $2k_{\rm F}$ is continuous (discontinuous) for odd (even) $J$.  
(e) and (f) show the intraband and interband polarizabilities, respectively. 
} 
\label{fig:pol_2DCG}
\end{figure}

Next we consider the polarizability for the doped extrinsic case (i.e., $n \neq 0$). At $q=0$, the polarizability becomes the DOS at Fermi energy $\varepsilon_{\rm F}$, i.e., $\Pi(0)=N(\varepsilon_{\rm F}) \equiv N_0$, where $N(\varepsilon)={g t_{\perp} \over 2\pi J (\hbar v_0)^2} \left({|\varepsilon|/t_{\perp}}\right)^{{2 \over J}-1}$ is the DOS at energy $\varepsilon$ for C2DES. 
Note that in the intrinsic limit ($n \rightarrow 0$, or equivalently $\varepsilon_{\rm F} \rightarrow 0$), $N_0$ diverges as $\varepsilon_{\rm F}^{{2\over J}-1}$ or $n^{1-{J \over 2}}$ for $J \ge 3$, which indicates enhanced screening at low densities. 
In Fig.~\ref{fig:pol_2DCG}(a) the Thomas-Fermi (TF) screening wave vector, $q_{\rm TF}=q_{\rm s}(q=0) = (2\pi e^2 /\kappa) \Pi(q=0)$, is shown as a
function of density, in which $q_{\rm TF}\propto N_0\propto n^{1-{J \over 2}}$ diverges as $n \rightarrow 0$ for $J \ge 3$.
For comparison, we also show the TF screening wave vector for
multilayer graphene in Fig.~\ref{fig:pol_2DCG}(b). At low densities,
the energy spectrum of multilayer graphene is described by a set of
C2DESs\cite{min2008}, and the C2DES with the largest chirality
dominates the DOS and the screening wave vector. Note that the periodic
ABC-stacked $J$-layer graphene, whose energy spectrum is described by
$J$ C2DES at low densities, shows divergent screening behavior for
$J\ge 3$.

Figure \ref{fig:pol_2DCG}(c) shows the total static polarizability as a function of wave vector for C2DES. We note that the normalized
polarizabilities of Fig.~\ref{fig:pol_2DCG}(c) vary in orders of
magnitude depending on $J$ if they are expressed in absolute units, as
indicated in Fig.~\ref{fig:pol_2DCG}(a). In Fig.~\ref{fig:pol_2DCG}(d)
the first derivative of the total polarizability with respect to
the wave vector, $d\Pi(q)/dq$, is shown for each $J$. In
Figs.~\ref{fig:pol_2DCG}(e) and \ref{fig:pol_2DCG}(f) we show the intraband and 
interband polarizabilities for several chirality indices $J=1,2,3,4,5$, respectively.   

For small $q$, the polarizability becomes
\begin{eqnarray}
\Pi_{\rm intra}(q) & = & N_0 \left [1-\frac{J^2 q^2}{8 k_{\rm F}^2} +O(q^4) \right ], \\
\Pi_{\rm inter}(q) & = & N_0 \frac{J^2 q^2}{8 k_{\rm F}^2} \left [ 1-\frac{J[(J-1)^2-5]}{16(J+2)} \frac{q^2}{k_{\rm F}^2} + O(q^4) \right ]. \nonumber
\end{eqnarray}
As $q$ increases, $\Pi_{\rm intra}$ ($\Pi_{\rm inter}$) decreases (increases) quadratically in the leading order and these quadratic leading order terms exactly cancel out in the total polarizability. Thus, the total polarizability increases slowly as $q^4$ except for $J=1$, where two terms exactly cancel out up to $q=2k_{\rm F}$, giving a constant 2D polarizability for $q \le 2k_{\rm F}$. 
For large $q$, we have $\Pi_{\rm intra}= N_0 \left[ (J/2)(q/k_{\rm F})^{-J} + O(q^{-J-1}) \right]$ and $\Pi_{\rm inter}=N_0 \left[ I(J)(q/k_{\rm F})^{2-J} + O(q^{-J}) \right]$. Thus for large wave vectors $q \gg 2k_{\rm F}$, the polarizability from interband transition dominates, and $\Pi(q) \approx \Pi_{\rm inter} (q)$.

Since many significant physical properties (e.g., transport, Kohn
anomaly, and quantum criticality) are induced by the behavior of the
polarizability at $q=2k_{\rm F}$, we look further into $\Pi(q)$ near
$q=2k_{\rm F}$. At $q=2 k_{\rm F}$, the calculated polarizability with
even $J$ shows a cusp while that with odd $J$ varies continuously with
$q$ appearing as an
inflection point in $\Pi(q)$. As $q$ approaches $2k_{\rm F}$ from
above we find $\Pi(q) \propto (q^2 -4k_{\rm F}^2)^{3/2}$ for an odd
number of $J$ and $ \Pi(q) \propto (q^2 - 4 k_{\rm F}^2)^{1/2}$ for
an even number of $J$. As a consequence, the static polarizability has a
discontinuous (continuous) first derivative for even (odd) $J$ [see
  Fig.~\ref{fig:pol_2DCG}.(d)]. For even $J$'s the first derivative is
discontinuous at $q=2k_{\rm F}$, showing $d\Pi({\bm q})/dq \propto
1/\sqrt{q^2-4 k_{\rm F}^2}$, while for odd $J$'s, the discontinuity
appears in the second derivative as  $d^2\Pi({\bm q})/dq^2 \propto
1/\sqrt{q^2-4 k_{\rm F}^2}$. The origin of these singular features is
closely related to $2 k_{\rm F}$ backscattering in the system. When
the contribution of the $2k_{\rm F}$ scattering to the polarizability
is enhanced due to the chiral property of even $J$ [or the overlap
factor in Eq.~(\ref{eq:bare_polarization})], the polarizability has a
singular feature, whereas when the backscattering is suppressed for
odd $J$, no singular features occur. Note that all these singular
properties arise only from the intraband polarizability.


So far we have studied the polarizability for a simple two-band model of 2D chiral quasiparticles. Now we consider the polarizability for a full-band continuum model of multilayer graphene near the hexagonal corners of the Brillouin zone, called the $K$ and $K'$ points, taking into account only nearest-neighbor intralayer hopping $t$ and the nearest-neighbor interlayer hopping $t_{\perp}$. 
In this paper
$t = 3$ eV and $t_{\perp} = 0.3$ eV will be used in numerical calculations.  

\begin{figure}
\includegraphics[width=0.8\linewidth,height=1.9in]{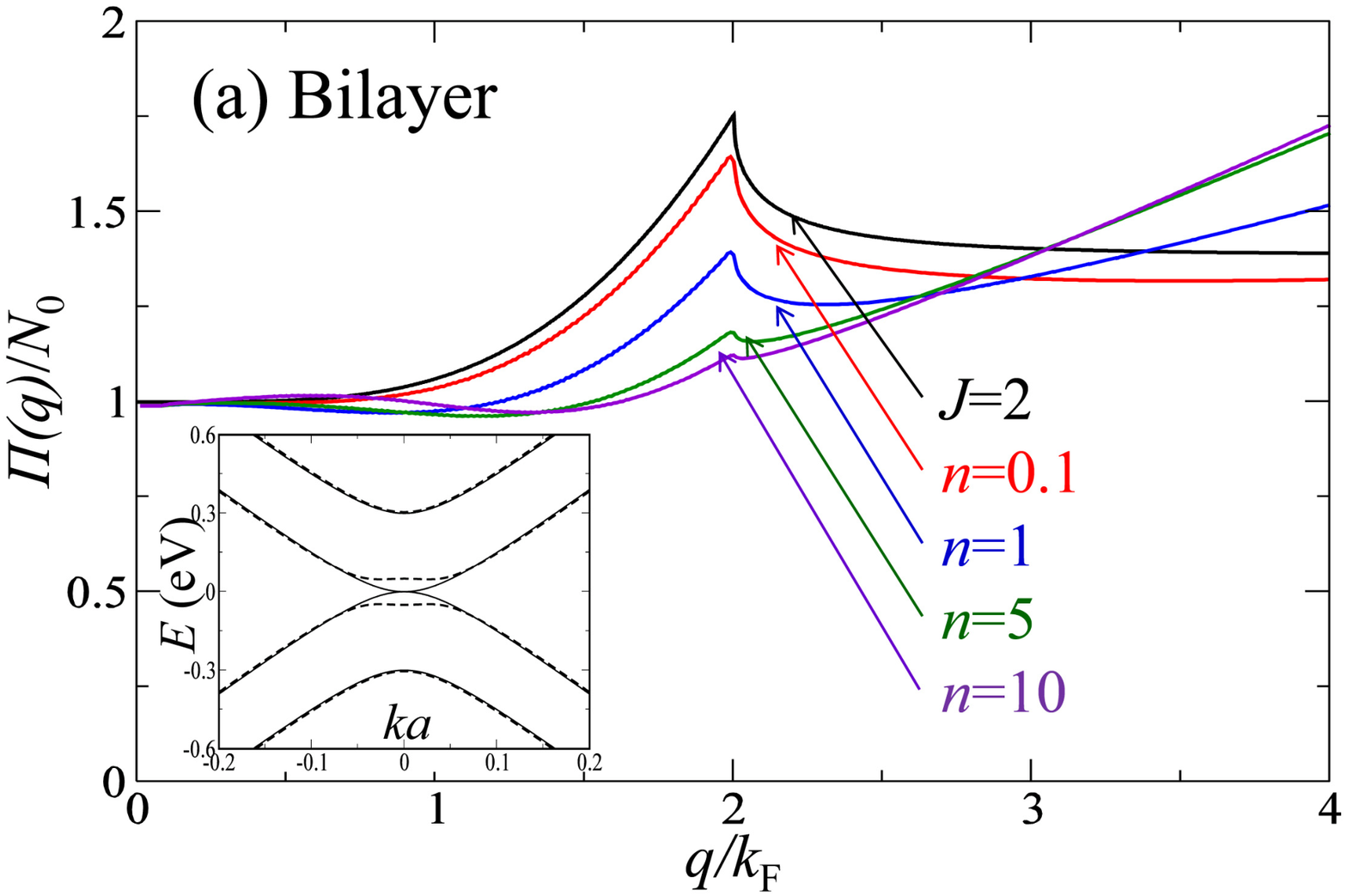} 
\includegraphics[width=0.8\linewidth,height=1.9in]{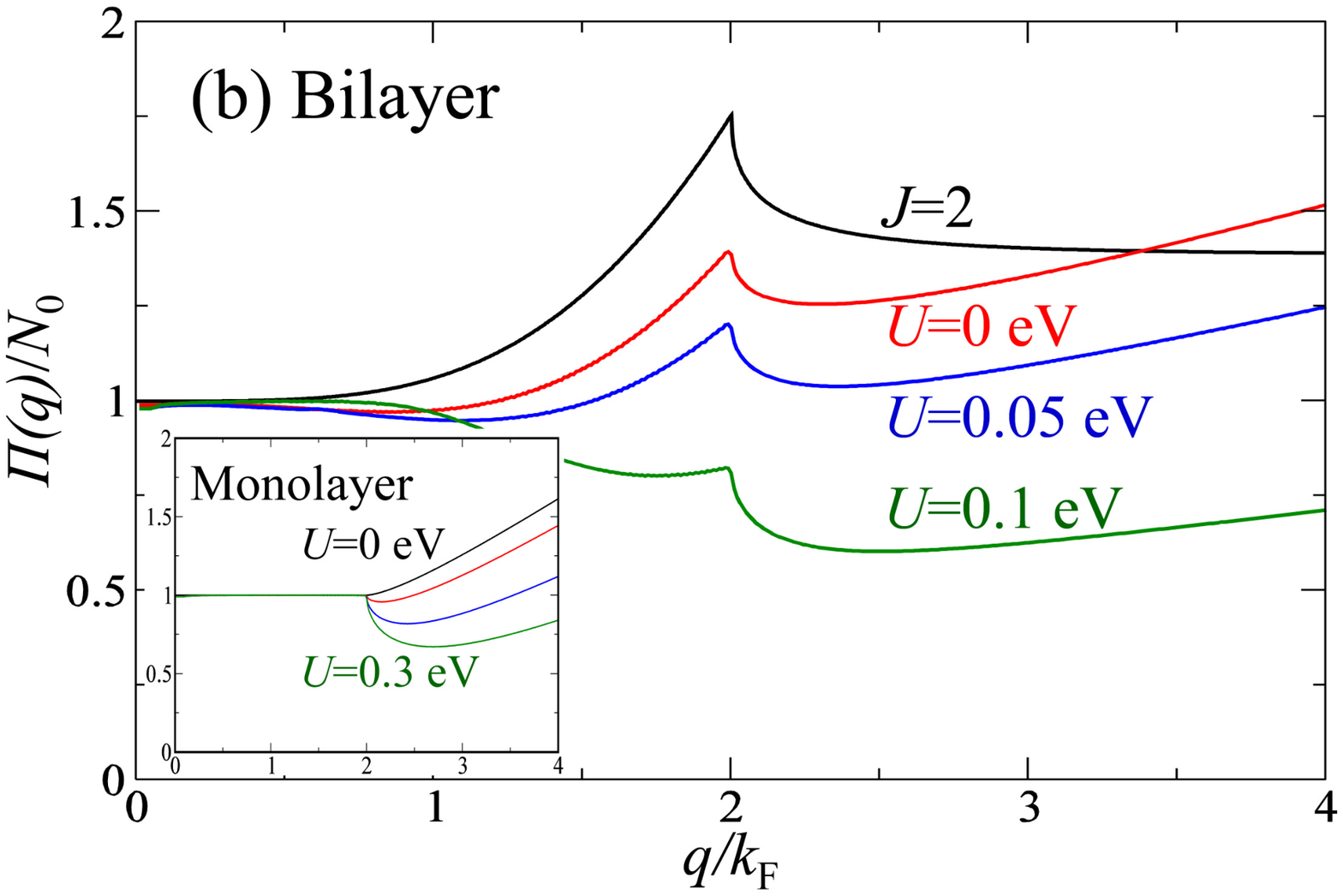}
\caption{(a) Static polarizability of bilayer graphene with the zero energy gap as a function of wave vector for several carrier densities in units of $10^{12}$ cm$^{-2}$. In (b) we include the on-site energy difference $U$ between the layers to open the energy gap. The polarizability is calculated for a carrier density $n=10^{12}$ cm$^{-2}$ and for different values of $U$. In both figures the black solid line corresponds to the result of C2DES with $J=2$. The inset of (a) shows the energy band structure of bilayer graphene for $U=0$ (solid line) and $U=0.1$ eV (dotted line). The inset of (b) shows the static polarizability of monolayer graphene with energy differences between the two sublattices of $U=0,0.1,0.2,0.3$ eV.} 
\label{fig:pol_bilayer}
\end{figure}

Figure \ref{fig:pol_bilayer}(a) shows the polarizability of zero-gap bilayer graphene as a function of wave vector for several carrier densities with the energy band structure as an inset. For small densities, the polarizability of the full four-band model resembles that of the $J=2$ C2DES (top black line). As the density increases, the polarizability eventually approaches that of monolayer graphene
because at high densities the interlayer  coupling becomes negligible and the energy band structure behaves as a collection of monolayer graphene sheets,
which is consistent with the previous results of Ref.~\onlinecite{screening_bilayer}.
The main difference between the polarizability of monolayer and high
density bilayer graphene is the existence of a kink at $q=2k_{\rm F}$,
which arises from the enhanced backscattering and does not vanish
completely in bilayer graphene even at very high densities, i.e., the
system maintains its memory even at very high carrier density where the
energy dispersion has become the same as that in monolayer
graphene.  

Figure ~\ref{fig:pol_bilayer}(b) shows the static polarizability of bilayer graphene with an on-site energy difference between the layers for a fixed density $n=10^{12}$ cm$^{-2}$. 
As the on-site energy difference increases, the energy band gap also increases, reducing the interband contribution and the overall magnitude of the polarizability. Note that in the presence of the on-site energy difference, the backscattering is suppressed and thus the cusp structure at $2 k_{\rm F}$ becomes weakened.
In the case of monolayer graphene, however, the on-site energy difference between the two {\it sublattices} gives rise to the enhanced backscattering, in addition to the opening of an energy gap. As a consequence, the polarizability of monolayer graphene with the on-site energy difference develops a cusp structure at $2 k_{\rm F}$, as shown in the inset of Fig.~\ref{fig:pol_bilayer}(b). This can be understood if we consider the effect of the on-site energy difference on the overlap factor in Eq.~(\ref{eq:bare_polarization}), 
which weakens the enhancement (suppression) of the backscattering for doped bilayer (monolayer) graphene.
Therefore, depending on the chiral properties of multilayer graphene, the on-site energy difference may enhance or weaken the backscattering and the cusp structure at $2k_{\rm F}$. 

\begin{figure}
\includegraphics[width=0.8\linewidth,height=1.9in]{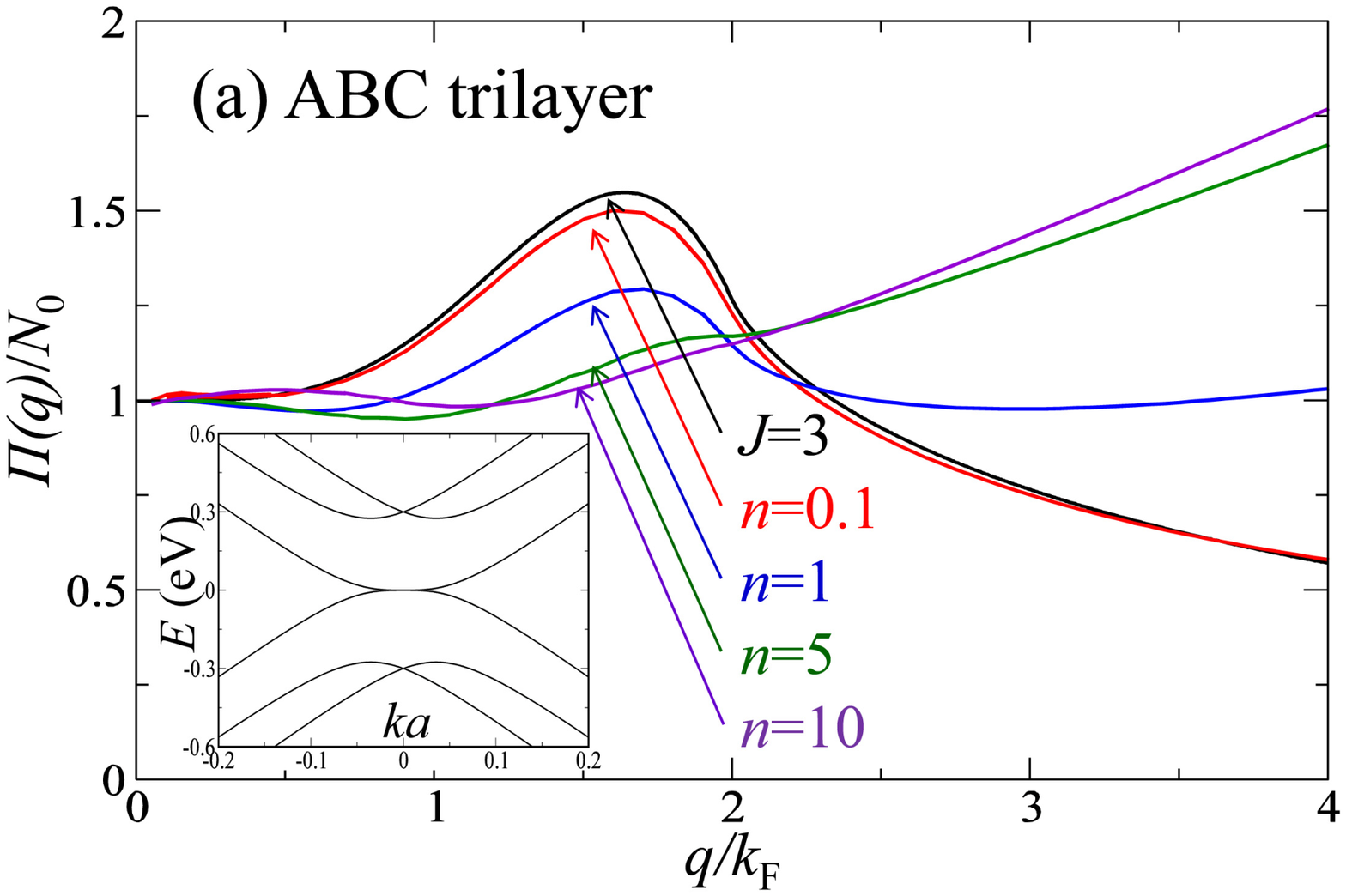}
\includegraphics[width=0.8\linewidth,height=1.9in]{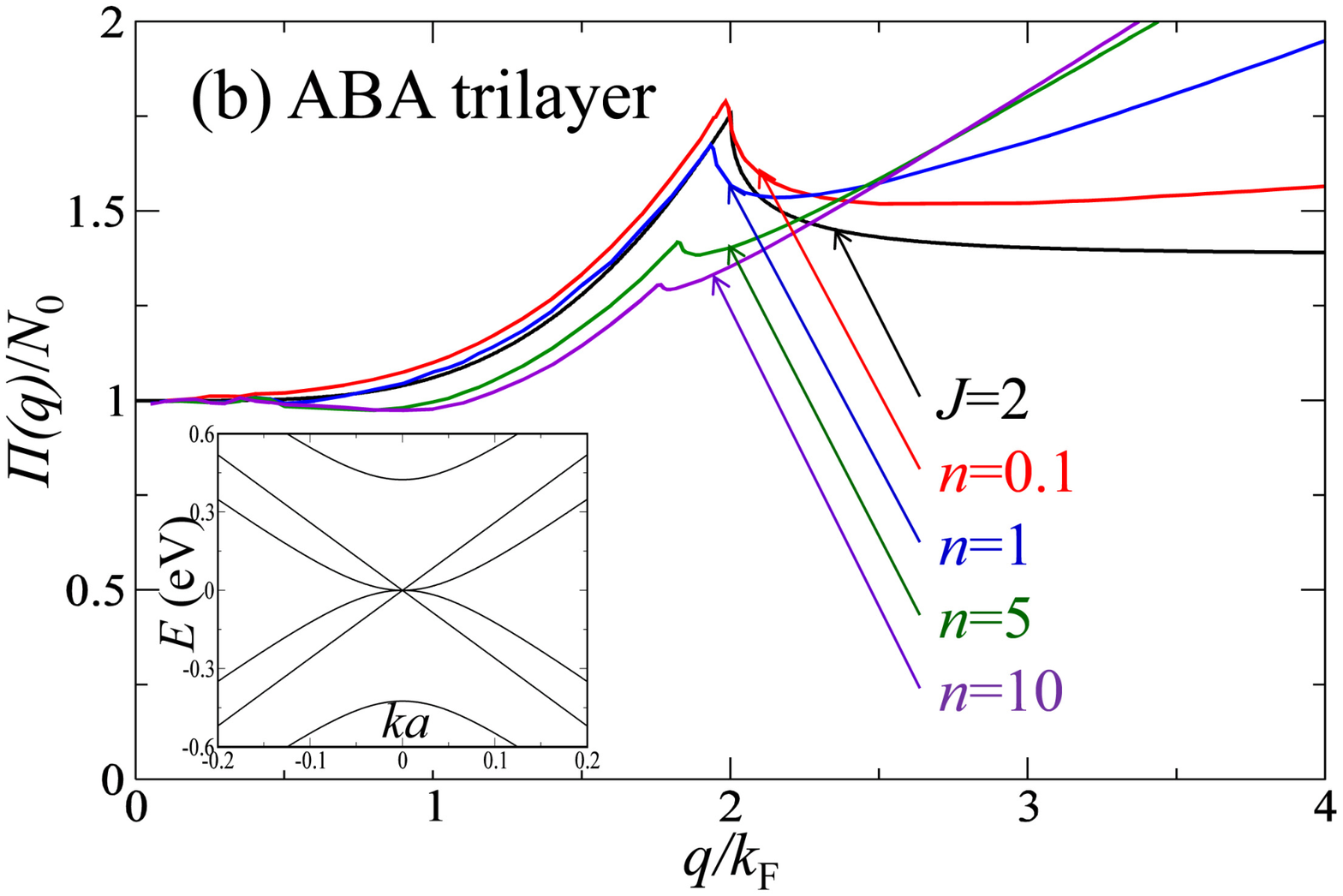}
\caption{Calculated polarizability as a function of wave vector with several densities in units of $10^{12}$ cm$^{-2}$ for (a) ABC trilayer graphene and (b) ABA trilayer graphene. The black line in (a) and (b) indicates the result of C2DES with $J=3$ and $J=2$, respectively. Insets show the energy band structure of each stacking.} 
\label{fig:pol_trilayer}
\end{figure}

In Fig.~\ref{fig:pol_trilayer} the polarizability of trilayer graphene is shown with the energy band structure as an inset. Excluding consecutive stacking sequences, there are two distinct stacking arrangements, ABC and ABA in trilayer graphene. Figure \ref{fig:pol_trilayer}(a) shows the polarizability of ABC trilayer graphene as a function of wave vector. Note that for ABC stacking the low-energy spectrum is described by $J=3$ C2DES. For comparison, we also show the polarizability of $J=3$ C2DES (top black line). At low densities, the polarizability resembles that of the low-energy C2DES with $J=3$ for ABC trilayer, while as the density increases, it approaches that of monolayer graphene. Figure \ref{fig:pol_trilayer}(b) is the same as Fig.~\ref{fig:pol_trilayer}(a) but for ABA trilayer graphene. Note that at low densities ABA trilayer graphene is described by a direct product of $J=1$ and $J=2$ C2DESs and the Fermi energy crosses two energy bands at $k_{\rm F_1}$ and $k_{\rm F_2}$ corresponding to the 
$J=1$ 
and $J=2$ energy bands, respectively. Thus the ABA trilayer graphene has two different Fermi wave vectors which can be calculated from the carrier density $n={g \over 4\pi} k_{\rm F}^2={g \over 4\pi} (k_{\rm F_1}^2+k_{\rm F_2}^2)$. The polarizability of ABA trilayer graphene in Fig.~\ref{fig:pol_trilayer}(b) appears to be qualitatively different from that of ABC trilayer graphene. One significant feature is the existence of a cusp at $2k_{\rm F_2}$ instead of $2k_{\rm F}$. 
As the density increases, the cusp becomes weakened and it shifts because the ratio $k_{\rm F_2}/k_{\rm F}$ decreases. This peculiar feature can be understood from the DOS of ABA trilayer graphene. Since the DOS of $J=2$ C2DES is larger than that of $J=1$ C2DES at small energies [i.e., $N(\varepsilon)\propto \varepsilon^{{2\over J}-1}$], many physical properties of ABA stacking trilayer graphene are determined by $J=2$ C2DES. Thus for low densities the polarizability resembles that of the $J=2$ C2DES and the cusp structure appears at $2 k_{\rm F_2}$. 

A similar analysis can be applied to arbitrarily stacked multilayer graphene and we can easily generalize the results of bilayer and trilayer graphene discussed so far. At low densities, the energy spectrum of multilayer graphene is described by a set of C2DESs and the polarizability behaves as that of a C2DES with the largest chirality. As the density increases, the polarizability follows that of monolayer graphene because the energy spectrum behaves as a collection of monolayer graphene sheets. In the presence of the on-site energy difference, the backscattering may be enhanced or suppressed depending on the chiral nature of the system, giving an enhanced or suppressed cusp structure at $2 k_{\rm F}$. 


\begin{figure}
\includegraphics[width=1\linewidth]{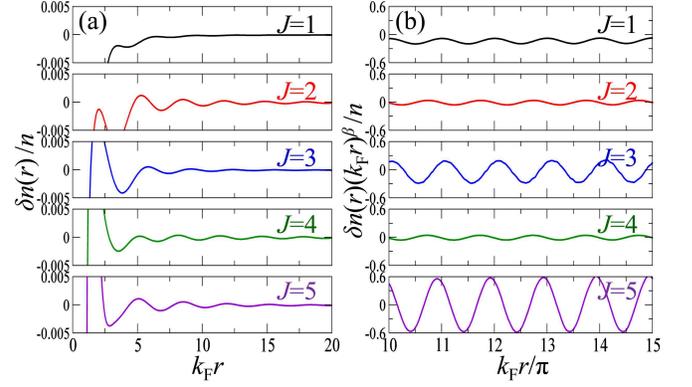} 
\caption{(a) Density oscillations of C2DES with chiralities $J=1,2,3,4,5$ using the RPA 
for $n=10^{12}$ cm$^{-2}$ with $\kappa=2.5$.
In (b) $\delta n(r)$ is rescaled by $r^{\beta}$, where $\beta=2$ ($\beta=3$) for even (odd) $J$, which clearly shows the decaying pattern at large distances with the periodicity of $\pi/k_{\rm F}$.
} 
\label{fig:pol_den_osc} 
\end{figure}

One direct consequence of the polarizability function we have calculated is the density fluctuation induced by a point charge, which is given by
\begin{equation}
\delta n(r)= \int {q dq \over 2\pi} J_0(qr)\left[\epsilon^{-1}(q)-1\right],
\end{equation}
where $J_0(x)$
is the zeroth-order Bessel function and $\epsilon(\bm{q})$ is the
static dielectric function, 
which is given by $\epsilon(\bm{q})=1+{2\pi e^2 \over \kappa q}\Pi(\bm{q})$ 
within the random phase approximation (RPA).  
Figure \ref{fig:pol_den_osc}(a) shows the calculated density oscillations of C2DES 
for $n=10^{12}$ cm$^{-2}$ with $\kappa=2.5$.
At a large distance ($k_{\rm F} r \gg 1$) we find a density oscillation
(Friedel oscillation) which has the form 
$\sin(2k_{\rm F} r)/r^2$ for even $J$ while $\cos(2k_{\rm F} r)/r^3$ for odd $J$,  
as seen in Fig.~\ref{fig:pol_den_osc}(b). This radial decaying pattern arises from the kink structure at $2k_{\rm F}$ of the polarizability function. When the discontinuity appears in the first (second) derivative of the polarizability, $d\Pi(q)/dq$ ($d^2\Pi(q)/dq^2$), the radial decay has a form of $r^{-2}$ ($r^{-3}$). The discontinuity of the derivative is the direct consequence of the enhanced (suppressed) backscattering for even (odd) $J$. 
One interesting aspect of the polarizability of C2DES, which is reflected in these Friedel oscillations as well as in the screening properties, is that even (odd) $J$ behaves as qualitatively similarly to the Lindhard polarization function for 2D (3D) electron gas systems.

 


In conclusion, we calculated the polarizability of multilayer graphene and studied the effect of stacking arrangement, carrier density, and on-site energy difference on graphene screening properties. We emphasize that knowing the screening function of multilayer graphene systems is crucial to understand the electronic properties of these systems because screening determines many fundamental properties, e.g., transport through screened Coulomb scattering by charged impurities, Kohn anomaly in phonon dispersion, Friedel oscillation of electron potential, and the RKKY interaction in the presence of magnetic impurities.

This work is supported by the NRI-SWAN and US-ONR.\\


\end{document}